\documentclass[onecolumn]{aastex631}
\usepackage{xcolor}
\usepackage{amsmath}
\usepackage{lipsum}
\usepackage{comment}
\usepackage{makecell}

\usepackage{hyperref}
\usepackage{csvsimple}

\usepackage{xcolor}            
\usepackage{ulem}                          

\newcommand{\skipthis}[1]{\relax}

\received{August 14, 2023}
\revised{September 29, 2023}
\accepted{October 4, 2023}

\submitjournal{The Astrophysical Journal Letters}

\shorttitle{Hyades Ultramassive White Dwarf}
\shortauthors{Miller et al.}

\begin{document}

\title{An Extremely Massive White Dwarf Escaped From the Hyades Star Cluster}

\correspondingauthor{David R. Miller}
\email{drmiller@phas.ubc.ca}

\author[0000-0002-4591-1903]{David R. Miller}
\affiliation{Department of Physics and Astronomy, University of British Columbia, Vancouver, BC V6T 1Z1, Canada}

\author[0000-0002-4770-5388]{Ilaria Caiazzo}
\affiliation{Division of Physics, Mathematics and Astronomy, California Institute of Technology, Pasadena, CA 91125, USA}

\author[0000-0001-9739-367X]{Jeremy Heyl}
\affiliation{Department of Physics and Astronomy, University of British Columbia, Vancouver, BC V6T 1Z1, Canada}

\author[0000-0001-9002-8178]{Harvey B. Richer}
\affiliation{Department of Physics and Astronomy, University of British Columbia, Vancouver, BC V6T 1Z1, Canada}

\author[0000-0002-6871-1752]{Kareem El-Badry}
\affiliation{Center for Astrophysics | Harvard \& Smithsonian, 60 Garden Street, Cambridge, MA 02138, USA}

\author[0000-0003-4189-9668]{Antonio C. Rodriguez}
\affiliation{Division of Physics, Mathematics and Astronomy, California Institute of Technology, Pasadena, CA 91125, USA}

\author[0000-0002-0853-3464]{Zachary P. Vanderbosch}
\affiliation{Division of Physics, Mathematics and Astronomy, California Institute of Technology, Pasadena, CA 91125, USA}

\author[0000-0002-2626-2872]{Jan van Roestel}
\affiliation{Division of Physics, Mathematics and Astronomy, California Institute of Technology, Pasadena, CA 91125, USA}
\affiliation{Anton Pannekoek Institute for Astronomy, University of Amsterdam, NL-1090 GE Amsterdam, the Netherlands}

\begin{abstract}

We searched the Gaia DR3 database for ultramassive white dwarfs with kinematics consistent with having escaped the nearby Hyades open cluster, identifying three such candidates. Two of these candidates have masses estimated from Gaia photometry of approximately 1.1 solar masses; their status as products of single stellar evolution that have escaped the cluster was deemed too questionable for immediate follow-up analysis. The remaining candidate has an expected mass $>1.3$ solar masses, significantly reducing the probability of it being an interloper. Analysis of follow-up Gemini GMOS spectroscopy for this source reveals a non-magnetized hydrogen atmosphere white dwarf with a mass and age consistent with having formed from a single star. Assuming a single-stellar evolution formation channel, we estimate a $97.8\%$ chance that the candidate is a true escapee from the Hyades. With a determined mass of 1.317 solar masses, this is potentially the most massive known single-evolution white dwarf and is by far the most massive with a strong association with an open cluster.

\end{abstract}



\section{Introduction}\label{sec:intro}

White dwarfs (WDs) form at the end of a low to intermediate-mass star's life. At this point in its evolution, the star has lost the ability to sustain ongoing fusion, shedding its outer layers and leaving behind its dense core. Outside of the degenerate core, which is typically composed of carbon-oxygen (CO) or oxygen-neon (ONe), thin layers of lighter elements make up its outer envelope. WDs are the expected final fate for more than $97\%$ of all stars in the Milky Way \citep{2001PASP..113..409F}. Though these compact objects no longer generate radiation via ongoing fusion, they radiate thermal energy from earlier such events, slowly cooling over time.

The Chandrasekhar limit, with a value around 1.39 M$_{\odot}$, provides the maximum mass of a stable WD \citep{1987ApJ...322..206N}. On the other hand, the maximum mass of the WD's precursor is far more contentious. Theoretical studies tend to favour a maximum mass around 8 M$_{\odot}$ \citep[e.g.][]{1983A&A...121...77W}, while observed supernova II rates, indicative of the death of a higher mass star on its way to becoming a neutron star (NS) or black hole (BH), point to a larger upper limit \citep{2011ApJ...738..154H}, potentially as high as 12 M$_{\odot}$ \citep{2003ApJ...598.1076K}. Better constraints on this maximum mass can be obtained by studying the relation between the mass of the progenitor stars and the final mass of the WDs (initial-final mass relation or IFMR)  \citep[e.g.][]{2018ApJ...866...21C,2018ApJ...860L..17E,2021ApJ...912..165R,2022ApJ...926..132H,2022ApJ...926L..24M}. Constraining the mass of a WD's progenitor is not an easy task. By studying an isolated white dwarf, we can determine its cooling age, i.e. the time elapsed since its formation as a white dwarf, but we have no indication on the total age of the star and, therefore, it is hard to constrain the mass of its progenitor. On the other hand, if a white dwarf is part of a star cluster, the total age of the star is the same as the cluster's, and by subtracting the cooling age from the total age we can infer the age at which the progenitor left the main sequence and therefore constrain its mass. 

While open clusters are a primary location to study the WD IFMR, extensive studies of open clusters suggest a significant deficiency of cluster WDs \citep[e.g][]{1977A&A....59..411W,1998ApJ...504L..91R,2001AJ....122.3239K,2003AJ....126.1402K,2007AJ....133.1490W,2021ApJ...912..165R}. One of the closest and most heavily studied clusters, the Hyades, was found to be missing approximately $75\%$ of its expected WDs \citep{1992AJ....104.1876W}. While more distant cluster WDs may have cooled beyond detectability thresholds, this is less important for nearby clusters such as the Hyades. Some of the missing WDs may be hidden in binary star systems. However, this is unlikely to account for such a high fraction of missing white dwarfs given the current understanding of WD binary fractions in the local neighbourhood \citep{2017A&A...602A..16T}. An alternative possibility is that many of the WDs may have left their clusters via dynamical interactions or aspherical mass loss during the asymptotic giant branch (AGB) phase of stellar evolution \citep[see][and references therein]{2003ApJ...595L..53F,2007MNRAS.382..915H,2009ApJ...695L..20F}. Recent work has examined this scenario by attempting to reconstruct nearby young open clusters and identifying WDs whose motion is consistent with past cluster membership \citep[e.g.][]{2021arXiv211004296H,2022ApJ...926..132H,2022ApJ...926L..24M}. These efforts have been successful in identifying several ultramassive ($M > 1.05$ M$_{\odot}$) high-probability escapee WDs. In this paper, we detail the extension of these methods to the nearby Hyades cluster. 

We describe our Hyades cluster member and escapee samples in Section~\ref{sec:sample}, analyze and discuss candidate white dwarf escapees in Section~\ref{sec:WDescapees}, re-examine Hyades cluster age determination in Section~\ref{sec:alt_cluster_age}, and summarize our findings in Section~\ref{sec:conclusions}. From this search, we identified three ultramassive WDs with kinematics consistent with past cluster membership that are young enough to have been born in the cluster. We find that two of these are not massive enough to confidently associate with the cluster, while the remaining one is a high-probability cluster escapee. From follow-up spectroscopy, we find this cluster escapee WD is consistent with having formed from single-stellar evolution. With a derived mass of $1.317^{+0.014}_{-0.018}$~M$_{\odot}$, it is perhaps the most massive single-stellar evolution WD known \citep[see][]{2018ApJ...861L..13G,2020ApJ...898...84K}.


\section{Sample}\label{sec:sample}


\subsection{Current Cluster}\label{sec:cluster}

\begin{figure}
\centering
    \includegraphics[width=0.8\textwidth,trim=0.5in 2.6in 0.2in 1in]{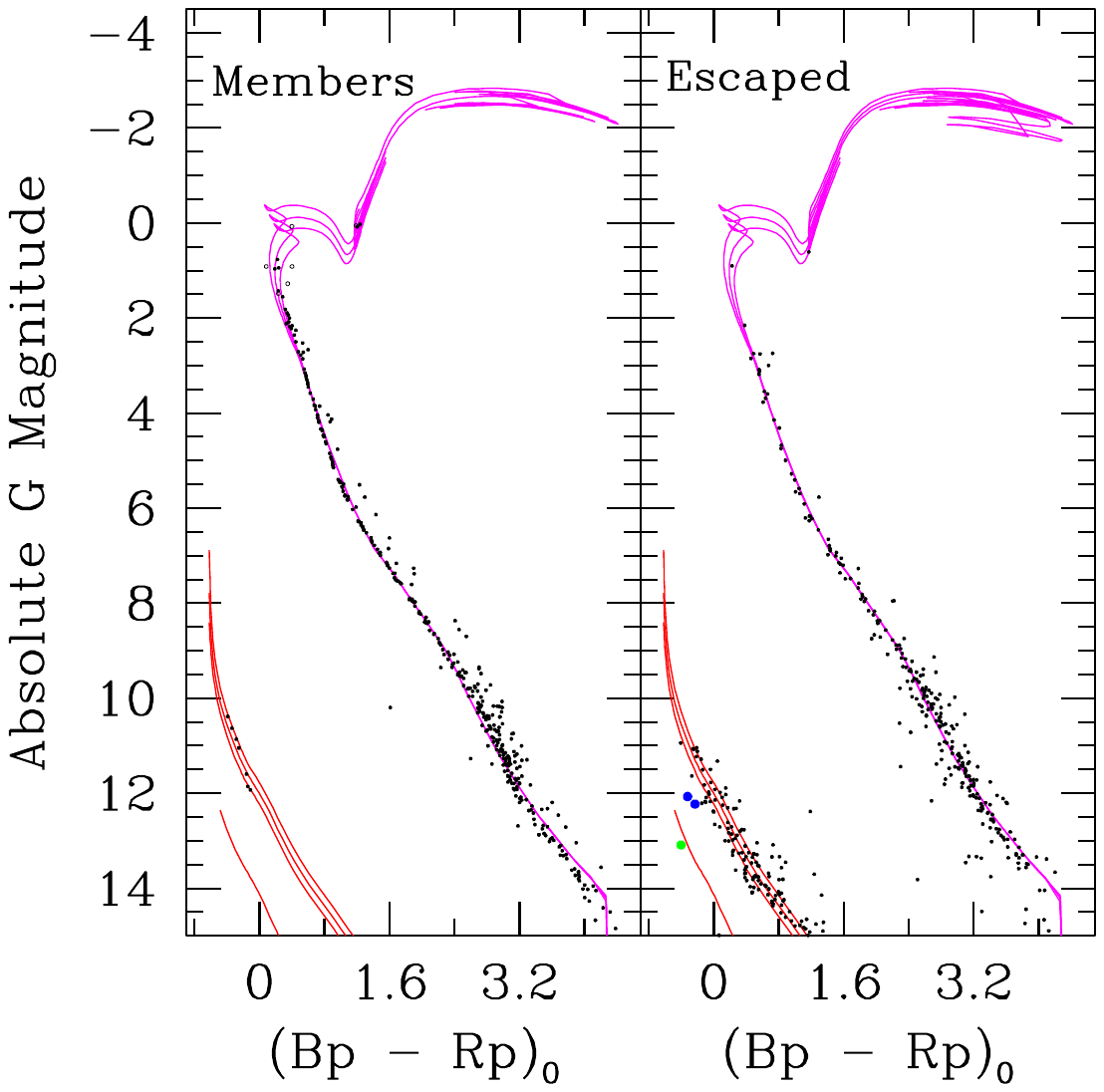}
    \caption{Hyades Gaia EDR3 color-magnitude diagram. Left: Current cluster {\bf members} together with Parsec \citep{2012MNRAS.427..127B} isochrones in purple ($600-800$~Myrs, left to right) for a slightly metal-rich cluster ($z=0.024$) with no reddening. Open circles are known variable or double stars \citep{1998A&A...331...81P}, and were not considered in deciding the age of the cluster which we estimate at $675\pm72$~Myrs. Carbon-oxygen WD cooling sequences \citep{2020ApJ...901...93B} shown in red are for $0.6-0.8$ and 1.3~M$_\odot$ (right to left). Right: The same but for stars kinematically consistent with having {\bf escaped} the Hyades cluster. Candidate ultramassive WD1 plotted in green,  WD2 and WD3 in blue.}
    \label{fig:Gaia_cmd}
\end{figure}

To develop a Gaia EDR3 Hyades catalogue of current cluster members, we started with the \cite{2019A&A...623A..35L} Gaia DR2 Hyades catalogue. While this catalogue contains candidates up to 70 degrees away from the cluster center, we focus on the subset within $9~$pc of the cluster, which they define as the tidal radius, and whose stars are very likely bound \citep{2011A&A...531A..92R,2019A&A...623A..35L}. Using this sample, we crossmatch with the Gaia EDR3 catalogue using the routine gaiaedr3.dr2$\_$neighborhood (available at the Gaia archive\footnote{https://gea.esac.esa.int/archive/}), which is a pre-computed crossmatch that accounts for Gaia EDR3 proper motions. In many cases, this routine tends to identify multiple matches. To account for this we exclude sources whose angular distance change between Gaia DR2 and EDR3 is $>500~$mas. This sample of 381 high-probability cluster members is shown in the left panel of Fig.~\ref{fig:Gaia_cmd}. The mass-weighted cluster center mean displacement from the Sun in galactic coordinates is 
\begin{equation}
        {\bf r}_{\rm{cluster}} = (-43.49 \pm 0.17, 0.55 \pm 0.16, -17.05 \pm 0.15)~{\rm{pc}}.
\label{eq:clust_disp}
\end{equation}
\noindent
Taking the subset of sample stars with measured radial velocities in Gaia DR3 \citep{2022arXiv220800211G} we determine the cluster center mean velocity with respect to the Sun to be 
\begin{equation}
      {\bf v}_{\rm{cluster}} = (-41.71 \pm 0.40 , -19.20 \pm 0.05, -1.06 \pm  0.19)~{\rm{km}}~\rm{s}^{-1}.
\label{eq:clust_vel}
\end{equation}

We estimate the cluster's age by fitting PARSEC \citep{2012MNRAS.427..127B} isochrones to the sample of selected cluster stars for a metallicity of $Z=0.024$ \citep{1998A&A...331...81P} and no reddening \citep{2006AJ....132.2453T}. In our fitting, we ignored several stars near the main sequence turnoff, known to be variable or binary stars \citep{1998A&A...331...81P}. Fig.~\ref{fig:Gaia_cmd} displays isochrones from $600-800$~Myrs; while upper main sequence stars seem to prefer a higher age estimate, the youngest isochrone best matches the giants. Collectively, we estimate a cluster age of $675\pm72$~Myrs. 

While we will use this as our best estimate of the cluster age, we emphasize that the Hyades cluster age determinations vary greatly in the literature, depending on the method and even when the same method is used. \cite{2023AJ....165..108B} similarly used PARSEC isochrones for a sample of Gaia EDR3 Hyades members, finding an age of $775\pm20$~Myr. They obtained an older age using the same isochrones because they ignored the four giant stars (see Fig.~\ref{fig:Gaia_cmd}) that surpass the nominal Gaia DR3 saturation limit, but are high-confidence members of the cluster. \cite{2016A&A...585A...7K} found that an age of $630$~Myr was a good fit to the data using both PISA and DARTMOUTH isochrones with convective core overshooting, while isochrones without convective core overshooting suggest a younger age of around $550$~Myr. \cite{2018ApJ...863...67G} estimated the cluster's age using models based on MESA stellar evolution code \citep{2019ApJS..243...10P,2018ApJS..234...34P,2015ApJS..220...15P,2013ApJS..208....4P,2011ApJS..192....3P}, obtaining best-fit ages between $589$ and $776$~Myr, depending on the photometry used and the degree of rotation (age estimates increase with rotation). \cite{2019A&A...623A..35L} found an age of $640^{+67}_{-49}$~Myr from the WD sequence using the IFMR from \cite{2018ApJ...860L..17E} along with PARSEC isochrones. \cite{2018ApJ...856...40M} used brown dwarf evolutionary models from \cite{2015A&A...577A..42B} to estimate an age of $650\pm75$~Myr from the lithium depletion boundary. \cite{2023MNRAS.518..662B} suggested an age of $635\pm135$ Myrs based on a compilation of eight literature age estimates.

Given the high degree of uncertainty in the cluster age, when vetting potential escapee candidates we allow for WDs with total age estimates within $2\sigma$ of our best cluster age estimate (from 531 to $819$~Myr). The high degree of uncertainty in the cluster age prevents determining reasonable estimates for the progenitor mass of any examined WD, as the progenitor lifetime will be far too uncertain for the highest mass WDs.

\subsection{Escapee Candidates}\label{sec:escapees}

While determining the cluster center and age confidently dictates a low degree of contamination, building an escapee sample prioritizes completeness over precision. To do this, we first redevelop a more complete sample of potential cluster members. Given the proximity of the Hyades cluster, for our initial cluster member candidate selection we queried the Gaia EDR3 archive for all sources with parallax $>5$~mas (that is, within 200~pc) and parallax over error $>10$, returning $2,234,316$ total sources. To be considered a potential cluster member we require the star to be within 15~pc of the determined cluster center in position space and within 30~mas~yr$^{-1}$ in proper motion space. This sample contains 521 stars. 

To identify potential Hyades escapees, we return to the complete sample of Gaia EDR3 sources with parallax $>5$~mas and parallax over error $>10$. To be considered a candidate escapee, a source has to exhibit a low proper motion relative to the cluster and to be moving in a manner that can place the star in the cluster within the cluster's lifetime. We employ the technique of \cite{2022ApJ...926..132H} to identify stars that meet these criteria, detailed below. We start by determining the distance of each source from the cluster center as a function of time, $d(t)$, for an arbitrary radial displacement $(\delta r)$
\begin{equation}
d(t)^2 = \left [ {\bf r}-{\bf r}_{\rm{cluster}} 
+  t \left ({\bf v}_{\rm{2D}}-{\bf v}_{\rm{cluster}} \right ) + \hat{\bf{r}} \delta r  \right ]^2,
\label{eq:4}
\end{equation}
where ${\bf v}_{\rm{2D}}$ is the velocity of the star in the plane of the sky and ${\bf r}$ is the star's displacement from the Sun, where we have assumed no relative acceleration. From this, we then determine the time of the star's closest approach to the cluster from 
\begin{equation}
t_{\rm{min}} = \frac{ \Delta {\bf r}\cdot  \Delta {\bf v} - \left (  \Delta {\bf r} \cdot \hat{\bf{r}}  \right ) \left (  \Delta {\bf v} \cdot \hat{\bf{r}}  \right ) }{\left (  \Delta {\bf v} \cdot \hat{\bf{r}}  \right )^2-\left (  \Delta {\bf v}\right )^2},
\label{eq:5}
\end{equation}
where 
\begin{equation}
\Delta {\bf r} = {\bf r}-{\bf r}_{\rm{cluster}} ~{\rm{and}}~
\Delta {\bf v} = {\bf v}_{\rm{2D}}-{\bf v}_{\rm{cluster}}.
\label{eq:6}
\end{equation}
From this, we estimate the radial velocity and displacement of the star from 
\begin{equation}
\delta r = v_r t_{\rm{min}} =  -\hat{\bf{r}} \cdot \left ( \Delta {\bf r} +t_{\rm{min}} \Delta {\bf v}   \right ) 
\label{eq:7}
\end{equation}
and are able to reconstruct the star's 3D velocity
\begin{equation}
\hat {\bf v}_{\rm{3D}} = {\bf v}_{\rm{2D}} + v_r \hat{\bf{r}}
\label{eq:8}
\end{equation}
along with its relative velocity
\begin{equation}
\Delta \hat {\bf v}_{\rm{3D}} = \hat {\bf v}_{\rm{3D}} - {\bf v}_{\rm{cluster}},
\label{eq:9}
\end{equation}
where the caret denotes a reconstructed quantity. 

We require that escapee candidates have a time of closest approach within the cluster's past lifetime that is within 30~pc of the cluster center. This distance cutoff was chosen as it compares to the cluster extent  \citep[see][]{2011A&A...531A..92R,2019A&A...623A..35L}. Additionally, we require a reconstructed 3D velocity relative to the cluster center less than 10~km~s$^{-1}$. These requirements collectively reduce the sample to $3,920$ candidate escapees, approximately $0.2\%$ of the original sample. The most massive stars in a cluster tend to gravitate towards the center, giving the most massive formed WDs an increased probability of a high-velocity ejection \citep{2013MNRAS.434.2509M}. Additionally, mass loss on the asymptotic giant branch increases significantly with initial mass \citep{2018A&ARv..26....1H}, increasing the strength of a potential natal velocity kick for a given degree of asymmetric mass loss. We allow such a large relative 3D velocity to account for these possibilities. However, we emphasize that main sequence stars and lower mass WD escapees are significantly less likely to be found with such high relative velocities. 

To avoid excessive contamination of main sequence stars, we reduce the maximum $\Delta \hat {\bf v}_{\rm{3D}}$ to 2~km~s$^{-1}$ for stars which we do not identify as likely WDs. This gives us our final sample of candidate escapees, which includes 145 likely WDs and 288 others, within 10 and 2~km~s$^{-1}$ of the cluster center motion, respectively. This combined sample of candidate escapees is shown in the right panel of Fig.~\ref{fig:Gaia_cmd}. Maintaining the high relative 3D velocity cutoff across the WD sequence leads to a high level of expected contamination at lower masses. The vast majority of these contaminating sources are eliminated as escapee candidates when considering source ages.


\section{White Dwarf Escapee Candidates }\label{sec:WDescapees}

\begin{figure}
    \centering
    \includegraphics[width=0.9\textwidth]{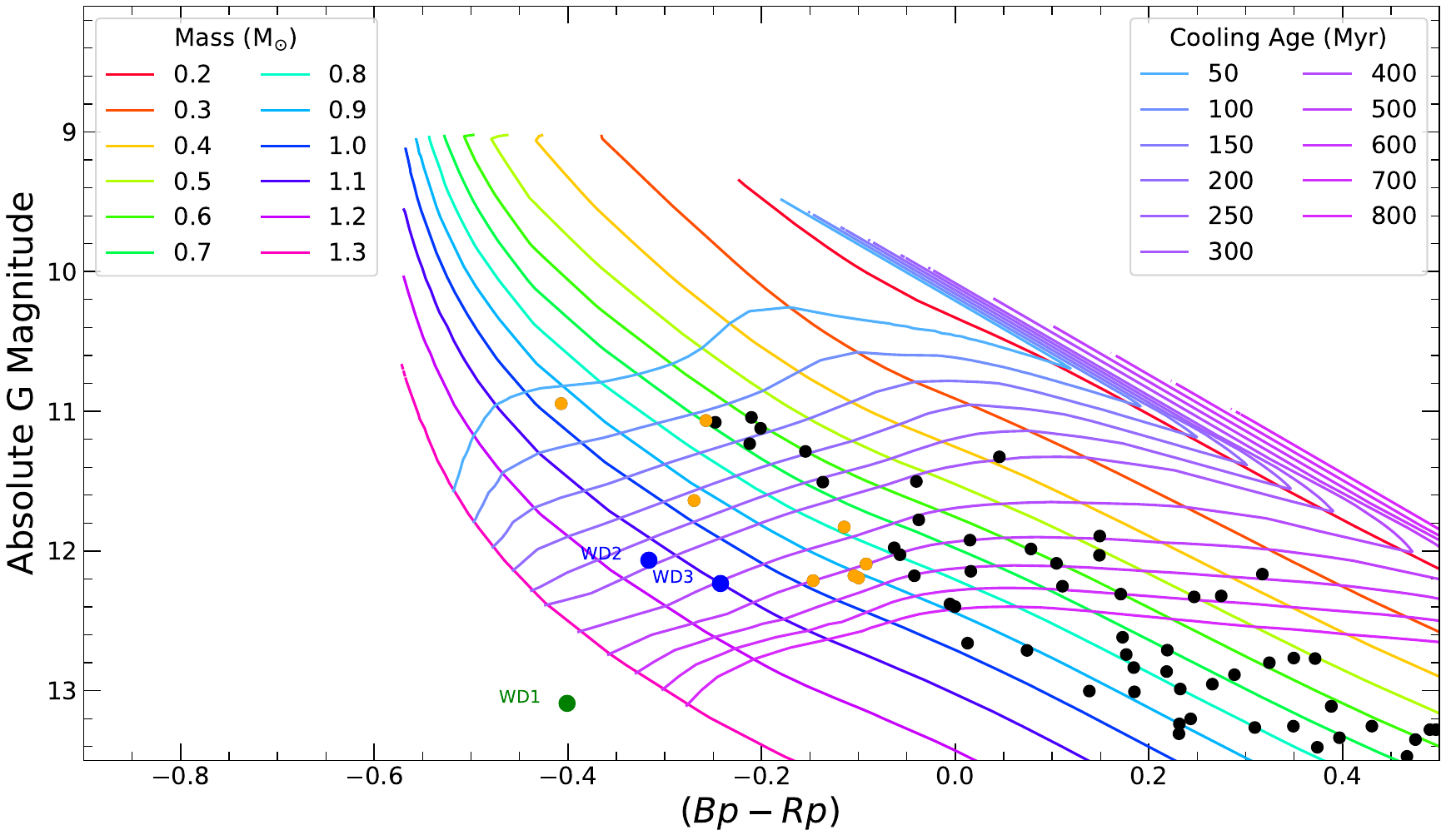}
    \caption{Candidate white dwarf escapees. Constant mass contours are shown from $0.2$ to $1.3$~M$_\odot$ (top right to bottom left), while age contours run horizontally for $50$ to $800$ Myrs (top to bottom). Candidates selected for follow-up are labeled WD1-WD3 (green and blue). Orange points have total ages consistent with potential past cluster membership but were not followed-up as their expected masses are below $1.0$~M$_\odot$. Black points are WDs with kinematics consistent with having escaped the Hyades but appear to be too old for potential past cluster membership.}
    \label{fig:escapee_cmd-wds}
\end{figure}

We estimate the mass and cooling age of the aforementioned sample of white dwarf escapee candidates using \cite{2020ApJ...901...93B} H atmosphere carbon-oxygen (CO) core cooling models, as displayed in Fig.~\ref{fig:escapee_cmd-wds}. In this sample, we identify three potential ultramassive WDs with cooling ages less than the cluster's age. Though we expect ultramassive WDs to have oxygen-neon (ONe) cores typically \citep{2007A&A...476..893S}, we focused on initial estimates for CO models for consistency across the WD sequence. This leads to an overestimation of the predicted mass compared with ONe models \citep[see Fig.~6 of][]{2022MNRAS.511.5198C}, particularly for WDs above 1.29~M$_\odot$, where general relativistic effects become significant \citep[see Fig.~7 of ][]{2022A&A...668A..58A}. That said, the difference is negligible at 1.05~M$_\odot$, so the candidate selection requirements for follow-up are not overly impacted. Additionally, mass estimates from Gaia photometry are somewhat uncertain due to sensitivity to assumed reddening (which is taken as zero here), so confident mass determination dictates follow-up with spectroscopy. 

The most massive of these three identified WDs, hereafter WD1 (Gaia EDR3 560883558756079616), has a mass estimate $>1.3$~M$_\odot$, this WD is also in the Sloan Digital Sky Survey \citep[SDSS,][]{2017arXiv171103234K} as SDSS J023836.27+764219.0. The remaining two, hereafter WD2 and WD3 (Gaia EDR3 3776918275016618112 and 3072348715677121280, receptively), each have mass estimates of approximately 1.1~M$_\odot$. WD2 was previously included in the \cite{1999ApJS..121....1M} catalogue as WD1043-050, while WD3 is in the SDSS footprint as SDSS J084214.98-022226.7. 

We estimate the progenitor mass of each WD in the escapee sample by interpolating the
initial-final mass relation built from \cite{2022ApJ...926L..24M} (for $M_{\rm f}>0.65$~M$_\odot$), and \cite{2018ApJ...866...21C} (for $M_{\rm f}<0.65$~M$_\odot$). From the determined progenitor mass, we estimate the progenitor lifetime using PARSEC isochrone tables at the metallicity of the Hyades ($Z=0.024$) with no reddening, giving a rough estimate of the total age of the WD when combined with the cooling age. From this, we reduce the sample of potential escapees to only those whose total age is below the upper 2$\sigma$ bound of the cluster's age. The significant age range allows us to account for both the uncertainty in the cluster's age and sensitivity of cooling age estimates due to the assumption of zero reddening. In addition to the three sources selected for potential follow-up, eight additional escapee candidates meet this criterion. Each has a mass estimated below 1.0~M$_\odot$; due to the focus of this work on ultramassive WDs, we did not consider  these sources further.

The remaining candidates are likely interlopers as their total ages are older than the cluster. Of the kinematically consistent likely WDs with masses estimated below 0.9~M$_\odot$, $68\%$ have relative 3D velocities of at least 6~km~s$^{-1}$. As previously mentioned, lower-mass WDs are less likely to be found with high relative 3D velocities and are inherently more likely to be interlopers. Additionally, the overall distribution of WDs reveals a narrow peak at around 0.6~M$_\odot$, with a notable pile-up between $0.7-0.9$~M$_\odot$, while WDs above 1.0~M$_\odot$ are much rarer \citep{2016MNRAS.461.2100T,2020ApJ...898...84K}. Because they are common, we expect a significant number of lower-mass WDs to be coincidentally consistent with past cluster membership, while this becomes much less likely with increasing mass.

\subsection{Interloper Contamination}\label{sec:interlopers}

We examine the expected degree of contamination using the \cite{2021MNRAS.508.3877G} EDR3 white dwarf catalogue (hereafter Fusillo). We search the Fusillo catalogue for high-probability WDs (their $\rm{Pwd}>0.9$) within 200~pc, with parallax over error $>10$, returning $72,071$ sources. We estimate each source's mass, cooling age, progenitor mass, and progenitor lifetime as before, finding that $1,606$ have total ages under 819~Myrs (the cluster's age plus 2$\sigma$), giving a fraction of $2.2\%$ of white dwarfs young and massive enough to be born in the Hyades. Of these younger sources, 317 ($19.7\%$) have mass estimates above 1.1~M$_\odot$, 50 ($3.1\%$) of which are above 1.3~M$_\odot$. 

Applying the same $\rm{Pwd}\geq0.9$ cut to our escapee catalogue, we find 131 candidate escapees, of which 11 are younger than 819~Myrs. This gives a youth percentage of $8.4\%$, well above expectations suggesting the presence of some true escapees. For a random sample of 131 escapee candidates, we predict just 2.92 to be younger than 819~Myr, of which 2.34, 0.58. and 0.09 are expected to be $<1.1$~M$_\odot$, $>1.1$~M$_\odot$, and $>1.3$~M$_\odot$. This gives us a crude estimate for the probability of identifying at least one single interloper in our sample above $1.1$~M$_\odot$ and $1.3$~M$_\odot$ as $44\%$ and $9\%$, respectively. Given such a high probability of a single interloper above $1.1$~M$_\odot$, it will be difficult to say with any confidence that WD2 and WD3 are both likely cluster escapees, especially given the high degree of uncertainty in the cluster's age. As such, we opt to focus our remaining work on the lone $>1.3$~M$_\odot$ WD, WD1. 

When ignoring total age, we find that just $0.25\%$ of WDs in this sample appear to be $\geq1.3~\rm{M}_{\odot}$. The mean heliocentric distance of this sample is 114.2~pc, more than twice the distance to the Hyades cluster center. This could bias results because more distant WDs are less likely to be detected at a given temperature. Restricting the sample to those within 70~pc, which gives a mean heliocentric distance comparable to the Hyades cluster center, only slightly shifts the proportion of $\geq1.3~\rm{M}_{\odot}$ to $0.28\%$. Given that this rate is only marginally higher than that found in the 200~pc sample, we do not find that the 200~pc distance cutoff overly biases statistics against high-mass WDs. 

The way the Hyades moves in space may be special compared to a random sample, leading to an inherent overabundance of high-mass escapee candidates. To examine this possibility, we first remove the restriction of WDs within 200~pc in the escapee catalogue. This adjusts the catalogue to include 154 candidate escapees, 13 of which qualify as potential escapees based on age. Neither of the additional young candidates is $>1.1$~M$_\odot$, and the interloper fraction remains at $91.6\%$. We then shift the cluster center in ten steps of 30~pc along the cluster's 3D trajectory and develop WD escapee catalogues for each redefined center. From these centers, we find a mean interloper percentage of $96.2\pm2.8\%$, with no candidate escapees $>1.3$~M$_\odot$. One potential issue is that the heliocentric distance to each cluster center varies. To consider whether this biases the above result, we maintain the cluster distance and instead move along its apparent path in the sky, again iterating ten times and tracing back escapee candidates. We find a mean interloper percentage of $94.2\pm1.2\%$, again identifying no candidate escapees $>1.3$~M$_\odot$. While this does suggest that maintaining the cluster center distance increases the number of potential true escapees, having identified no $>1.3$~M$_\odot$ WD escapees from any of the twenty shifted cluster centers provides strong support for WD1 being a true escapee from the Hyades cluster.

While we allow for any WD with a total age estimate younger than the cluster age plus 2$\sigma$ to be a potential escapee, those significantly younger than the cluster require cooling delays from past merger history to be escaped cluster members. Given this, we can provide an additional constraint on the probability of WD1 being an interloper, given that its age is consistent with that of a WD born from a single progenitor star in the cluster. Of the 50 WDs above $1.3$~M$_\odot$ with total ages younger than 819~Myrs in the previously examined 200~pc Fusillo catalogue sample, 27 are above the 2$\sigma$ lower bound cluster age of 531~Myrs. \cite{2023MNRAS.518.2341K} estimated that $56\%$ of $\simeq 1.3$~M$_\odot$ WDs formed via merger, allowing us to roughly estimate that approximately 12 of the 27 $>1.3$~M$_\odot$ WDs within 2$\sigma$ of the cluster age formed via single-stellar evolution. From this, we estimate the probability of a single-stellar evolution WD $>1.3$~M$_\odot$ within 2$\sigma$ of the cluster's age being coincidentally identified as a cluster escapee as just $2.2\%$. While we cannot entirely rule out the possibility that WD1 is an interloper, coincidental encounters with $>1.3$~M$_\odot$ WDs are rare enough to consider WD1 a high-probability Hyades escapee. 


\subsection{Spectroscopic Analysis of WD1}\label{sec:spectra}

We obtained follow-up spectroscopy for WD1 using the Gemini Multi-Object Spectrograph (GMOS) on the Gemini-North telescope \citep{2004PASP..116..425H,2016SPIE.9908E..2SG}. GMOS was set in long-slit mode, with a 1.0~arcsecond focal plane mask, the B600 grating, and no filter, centered at 520~nm. Data were binned 2x2 in spatial and spectral directions for an after-binning resolution of $\approx$1 angstrom. The total exposure time was 8,000 seconds, but three 1,000-second exposures proved unusable due to a bright contaminating star in the slit. The analysis below is based on the remaining five 1,000-second exposures. The spectrum shows broad and deep Balmer absorption lines indicative of a hydrogen atmosphere (DA) WD. We find no spectroscopic evidence of a significant magnetic field or fast rotation (both signs of past merger history, see \cite{1997MNRAS.292..205F,2012ApJ...749...25G,2013ApJ...773..136J,2020MNRAS.499L..21P}), nor does it have an excessively large transverse velocity, which can be an additional sign of past merger history \citep{2023MNRAS.518.2341K}. Collectively, these factors suggest WD1 likely formed via single-stellar evolution. 

\begin{figure}
    \centering
    \includegraphics[width=0.294\textwidth]{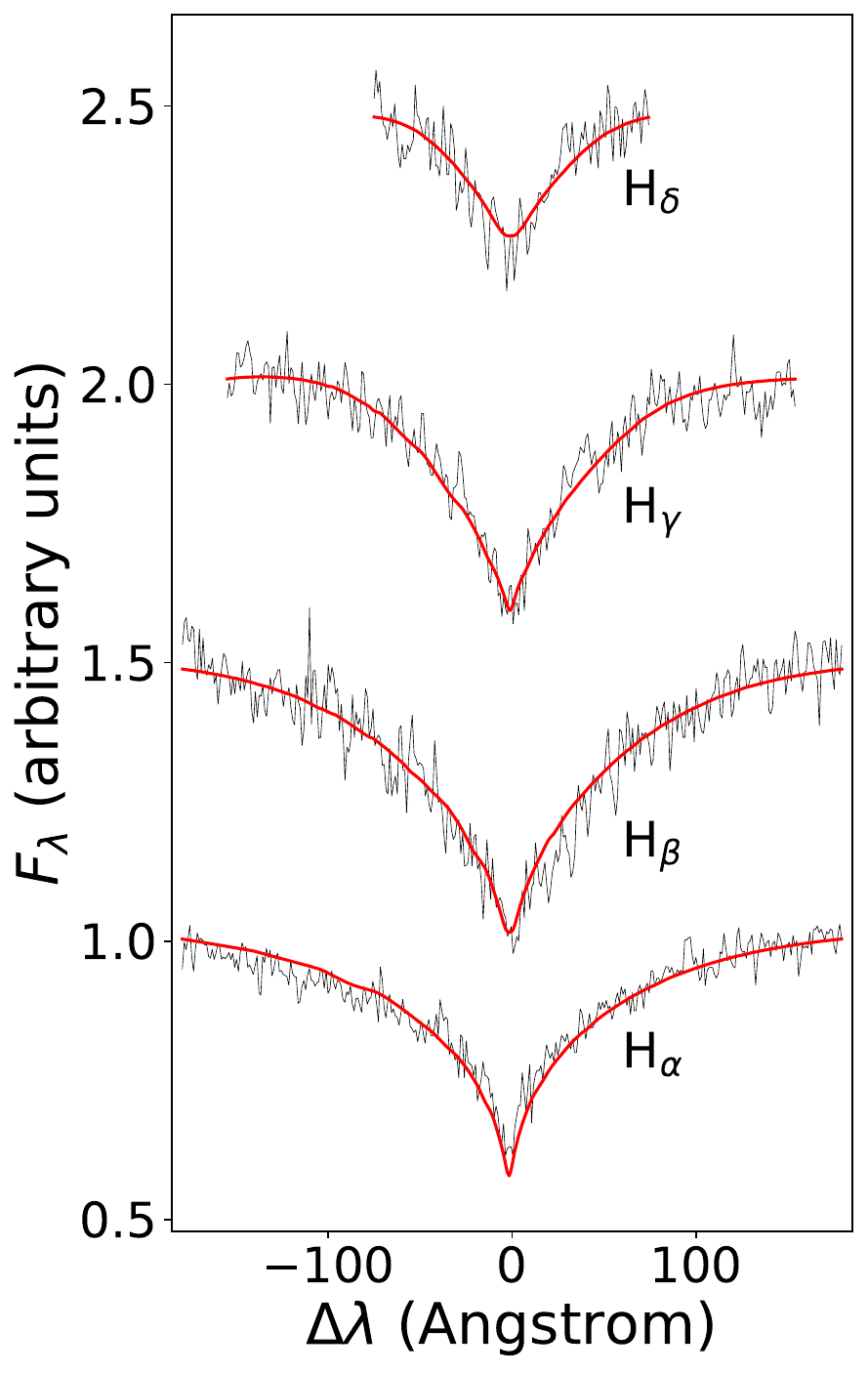}
    \includegraphics[width=0.692\textwidth]{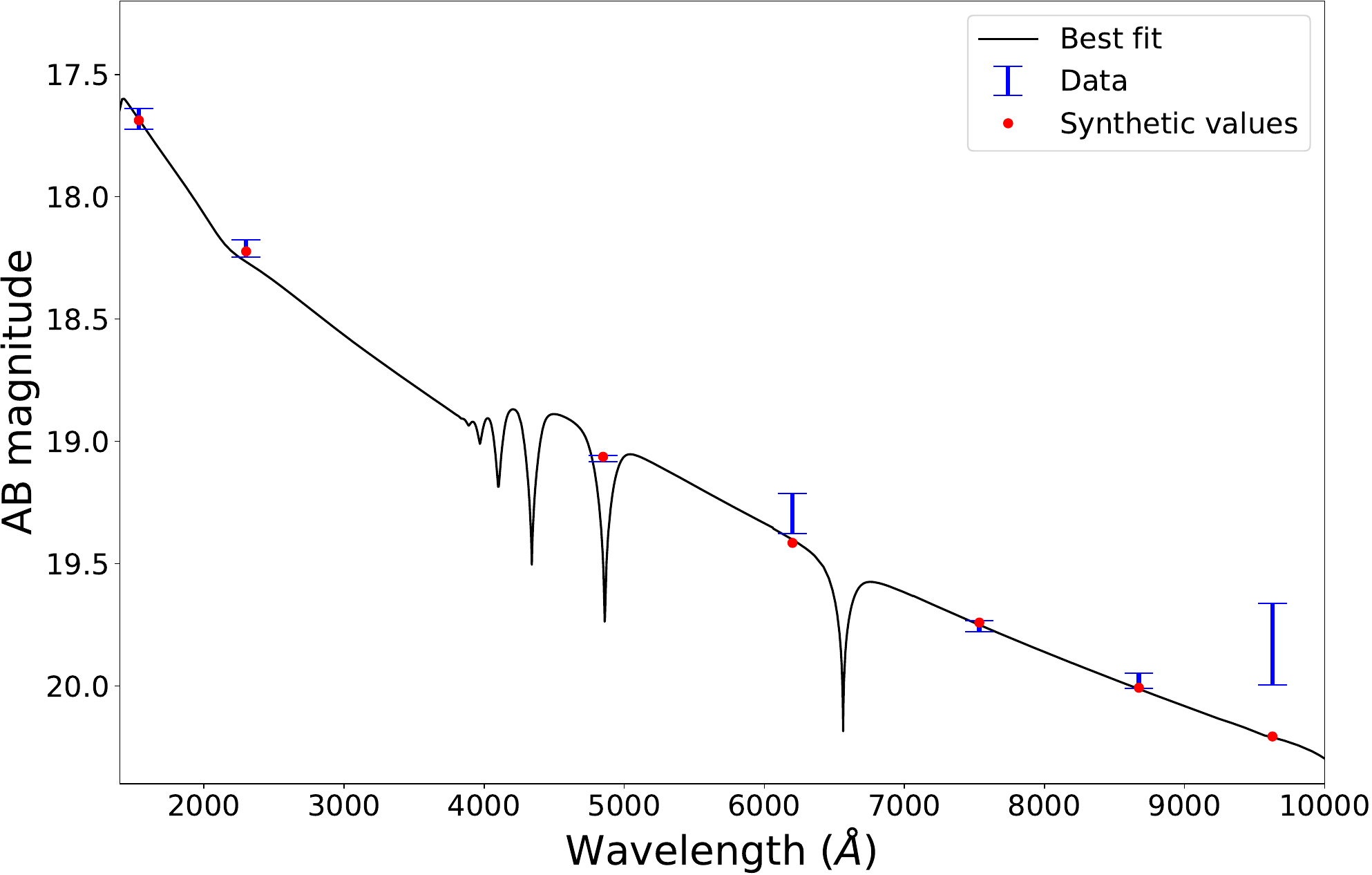}
    \caption{\textit{Left}: Gemini-North GMOS spectrum for WD1 isolating the first four Balmer lines (black). DA atmospheric model \citep{2011ApJ...730..128T} fit superimposed in solid red. \textit{Right}: Photometric fit of WD1 using available Galex and Pan-STARRS photometry with $1\sigma$ error bars (blue). Black lines shows the synthetic spectrum computed from the best-fit, with synthetic photometry shown in red. }
    \label{fig:WD1_spect}
\end{figure}

To determine atmospheric parameters, we employ non-local thermodynamic equilibrium (NTLE) pure DA models from \cite{2011ApJ...730..128T} extended to $\log (g/\mathrm{cm ~s}^{-2})= 10$. From these models, we determine the effective temperature ($T_{\rm eff}$), and $\log g$\, and the velocity from the redshift of the spectral lines ($v_z$) using a similar routine as \cite{2005ApJS..156...47L}: We start by fitting the spectrum to a grid of models blended with a polynomial in wavelength up to 9th order to deal with continuum calibration errors. The spectrum is then normalized by using points at a fixed distance from the Balmer lines. Finally, the Balmer lines are fit using the model spectra, returning the best-fit values of $T_{\rm eff}$, and $\log g$, and $v_z$. We use the Levenberg-Marquardt non-linear least squares minimization method in our fitting routines. The best simultaneous fit to the first four Balmer lines (Fig.~\ref{fig:WD1_spect}) gives $\log  (g/\mathrm{cm ~s}^{-2})=9.55\pm0.11$ and, $T_{\rm eff} = 26,400\pm200$~K, and $v_z=280\pm17$~km~s$^{-1}$. 

We attribute the measured redshift to two factors; Doppler shift due to the motion of the source away from us and gravitational redshift due to the strength of the source's gravity. We previously estimated a radial velocity of 20.4~km~s$^{-1}$ from escapee analysis, leaving the remaining $v_{z,g}=260$~km~s$^{-1}$ as the shift resulting from the gravitational redshift. To determine the uncertainty in the gravitational redshift estimate, we first note two sources of potential systematic error: radial velocity uncertainty and uncertainty in the best-fit $v_z$ as a result of the wavelength calibration. While our methods do not allow for a reliable estimate of the radial velocity uncertainty, the radial velocity may only differ by a few km~s$^{-1}$ for its reconstructed 3D velocity to remain consistent with being a cluster escapee. Systematic uncertainty in the best-fit value of $v_z$ primarily occurs due to instrument flexture during the $\approx$10 hours between the science and wavelength calibration exposures. Examination of the skylines in the wavelength calibration suggest a systematic uncertainty as high as 12~km~s$^{-1}$. Combining the statistical and systematic error, we estimate a velocity from gravitational redshift of $v_{z,g}=260\pm21$~km~s$^{-1}$.

While the spectrum of a WD gives insight into its atmospheric composition, the core is not directly discernible due to the opacity of the outer layers \citep{2008ApJ...683..978D}. WDs above about 1.05~M$_\odot$ are expected to harbor ONe cores, formed when core conditions in the late stages of stellar evolution are sufficient for off-center carbon ignition \citep{2014MNRAS.440.1274C}. While potential channels exist for WDs with CO cores to enter the ultramassive regime without igniting carbon \citep{2022MNRAS.511.5198C,2022MNRAS.512.2972W}, these channels are unlikely to maintain CO cores to the mass range of WD1. Using the determined $\log g$ and $T_{\rm eff}$, we estimate the mass and cooling age using \cite{2022A&A...668A..58A} ONe core WD cooling models, chosen due to their incorporation of full general relativistic effects on WD structure in their model evolution, which are thought to be particularly impactful for WDs above 1.29 M$_{\odot}$ \citep{2022A&A...668A..58A}. From linear interpolation of model tables, we find best-fit parameters of $M=1.317^{+0.014}_{-0.018}$~M$_\odot$, $R=2,230^{+280}_{-250}$~km, with $t_{\rm cool}=556^{+15}_{-22}$~Myrs. We summarize the astrometric, spectroscopic, and derived properties of WD1 in Table~\ref{tab:wd_derived}. 

The stated uncertainties are statistical uncertainties based on the best fit to the models, but additional systematic uncertainty may also be significant, particularly in the case of the cooling age. In the \cite{2019A&A...625A..87C} ONe evolutionary models, which are the basis for the relativistic extension given by \cite{2022A&A...668A..58A}, the authors consider a hydrogen envelope of $10^{-6}$~M$_\odot$. While a reduced hydrogen envelope should not overly impact the cooling age, an increase could lead to residual hydrogen burning, producing additional thermal radiation and delaying the WD's cooling. That said, WDs that retain a thicker H envelope are expected to be low mass WDs whose progenitors are low metallicity and also avoided the third dredge-up \citep{2021NatAs...5.1170C,2015A&A...576A...9A,2013ApJ...775L..22M}. Given the high mass of WD1, it is unlikely that residual hydrogen burning is a concern. Electron capture could also produce additional heat and delay the WD cooling. Still, this effect is likely only significant for WDs very close to the Chandrasekhar limit (see discussion in \cite{2020PhRvD.102h3031H}), thus not important for WD1.

Another potential concern is the possibility of a cooling delay due to phase separation. A significant delay that leads to an overabundance of WDs in a particular region of the HR diagram known as the Q branch \citep{2019ApJ...886..100C} has been studied in recent years. This appears to be primarily due to $^{22}\rm{Ne}$ phase separation \citep{2021ApJ...911L...5B}, which has been demonstrated to be important for ultramassive CO WDs but not likely for ONe core WDs \citep{2022MNRAS.511.5198C,2021A&A...649L...7C}, as is the expected core composition of WD1. Analogously, a similar process may come from the distillation of $^{56}\rm{Fe}$, which recent work suggests could delay cooling of high mass WDs by of order $100$~Myr  \citep{2021ApJ...919L..12C,2023ApJ...946...78C}. While this is certainly something to consider for ultramassive ONe WDs, further studies are required to understand better when and if this effect is critical to consider in WD cooling models. Though there are several potential cooling delays to consider, we do not currently find any that dictate imposing additional systematic uncertainty onto our stated statistical uncertainty.

As previously discussed, for such a high mass WD the degree of uncertainty in the cluster's age inhibits obtaining a well-constrained estimate of the corresponding progenitor lifetime, and thus, progenitor mass. That said, given our current understanding of the WD initial-final mass relation (IFMR) \citep{2018ApJ...866...21C,2018ApJ...860L..17E,2021ApJ...912..165R,2022ApJ...926..132H,2022ApJ...926L..24M}, we expect that such a high-mass WD would have evolved from a progenitor star of $M>7.5$~M$_\odot$ with a lifetime of $<40$~Myrs. For WD1 to be a single-stellar evolution WD that escaped from the Hyades cluster, the cluster age would have to be near the lower end of our age estimate of $675\pm72$~Myrs. Given the high probability that WD1 is a true escapee from the Hyades cluster, this places an additional constraint on the cluster's age.

A handful of white dwarfs thought to be smaller and more massive than WD1 have been observed, but all appear to be merger products because of their strong magnetic fields or rapid rotation \citep{1995MNRAS.277..971B,2020MNRAS.499L..21P,2021Natur.595...39C}. While these objects are certainly very massive, the strong fields make it difficult to constrain the masses firmly through spectroscopy, which is not an issue with WD1. As such, WD1 may be the most massive white dwarf with a well-defined mass determination and the smallest and most massive known white dwarf with a likely single star progenitor \citep{2018ApJ...861L..13G,2020ApJ...898...84K}. 

\begin{table}
    \caption{WD1 astrometric and photometric quantities from Gaia EDR3 (top row), additional photometry from Galex and Pan-STARRS1 (middle row), along with derived and spectroscopic quantities (bottom row). Spectroscopic estimates for radius, mass and age obtained with the assumption of an ONe core composition.}\label{tab:wd_derived}
    \centering
    \hskip-2.8cm\begin{tabular}{cccccccccc}
    \hline
    \multicolumn{1}{c}{$\rm{Gaia}$ $\rm{DR3}$ $\rm{Source}$ $\rm{ID}$} &
    \multicolumn{1}{c}{$\rm{RA}$} &  
    \multicolumn{1}{c}{$\rm{Dec}$} & 
    \multicolumn{1}{c}{$\rm{Parallax}$} & 
    \multicolumn{1}{c}{$\rm{pm~RA}$} & 
    \multicolumn{1}{c}{$\rm{pm~Dec}$} & 
    \multicolumn{1}{c}{$G$} & 
    \multicolumn{1}{c}{$\rm{Abs}$ $G$} &  
    \multicolumn{1}{c}{$B_p-R_p$} 
    \\
    &
    \multicolumn{1}{c}{$\rm{[deg]}$} &  
    \multicolumn{1}{c}{$\rm{[deg]}$} &  
    \multicolumn{1}{c}{$\rm{[mas]}$} &
    \multicolumn{1}{c}{$\rm{[mas~yr^{-1}]}$} &   
    \multicolumn{1}{c}{$\rm{[mas~yr^{-1}]}$} &   
    \multicolumn{1}{c}{$\rm{[mag]}$} & 
    \multicolumn{1}{c}{$\rm{[mag]}$} & 
    \multicolumn{1}{c}{$\rm{[mag]}$} 
    \\
    \hline  
    $560883558756079616$ &  $39.6517$ & $76.7052$ & $5.98$ & $46.46$ & $-26.49$ & $19.21$ & $13.09$ & $-0.40$ \\
    \end{tabular}
    \hskip-2.8cm\begin{tabular}{ccccccc}
    \hline
    \multicolumn{1}{c}{$\rm{PS1-}y$} &
    \multicolumn{1}{c}{$\rm{PS1-}z$} &  
    \multicolumn{1}{c}{$\rm{PS1-}i$} &
    \multicolumn{1}{c}{$\rm{PS1-}r$} &  
    \multicolumn{1}{c}{$\rm{PS1-}g$} &
    \multicolumn{1}{c}{$\rm{Galex-}NUV$} &  
    \multicolumn{1}{c}{$\rm{Galex-}FUV$}
    \\
    \multicolumn{1}{c}{$\rm{[mag]}$} &
    \multicolumn{1}{c}{$\rm{[mag]}$} &  
    \multicolumn{1}{c}{$\rm{[mag]}$} &
    \multicolumn{1}{c}{$\rm{[mag]}$} &  
    \multicolumn{1}{c}{$\rm{[mag]}$} &
    \multicolumn{1}{c}{$\rm{[mag]}$} &  
    \multicolumn{1}{c}{$\rm{[mag]}$}
    \\
    \hline  
    $19.83\pm0.17$ & $19.98\pm0.03$ & $19.76\pm0.02$ & $19.30\pm0.08$ & $19.071\pm0.013$ & $18.21\pm0.04$ & $17.68\pm0.04$ \\
    \end{tabular}
    \begin{tabular}{ccccccccccc}
    \hline
    \multicolumn{1}{c}{$d_{\rm{present}}$} & 
    \multicolumn{1}{c}{$v_r$} & 
    \multicolumn{1}{c}{$d_{\rm{min}}$} & 
    \multicolumn{1}{c}{$\Delta v_{\rm{3D}}$} &  
    \multicolumn{1}{c}{$t_{\rm{escape}}$} &
    \multicolumn{1}{c}{$\log g$} &
    \multicolumn{1}{c}{$T_{\rm{eff}}$} &
    \multicolumn{1}{c}{$v_{z,g}$} &
    \multicolumn{1}{c}{$R$} &
    \multicolumn{1}{c}{$M$} & 
    \multicolumn{1}{c}{$t_{\rm{cool}}$}
    \\
    \multicolumn{1}{c}{$\rm{[pc]}$} & 
    \multicolumn{1}{c}{$\rm{[km~s}^{-1}]$} & 
    \multicolumn{1}{c}{$\rm{[pc]}$} & 
    \multicolumn{1}{c}{$\rm{[km~s}^{-1}]$} & 
    \multicolumn{1}{c}{$\rm{[Myr]}$} &
    \multicolumn{1}{c}{$\rm{[cm}$ $\rm{s^{-2}]}$} &
    \multicolumn{1}{c}{$\rm{[K]}$} & 
    \multicolumn{1}{c}{$\rm{[km~s}^{-1}]$} & 
    \multicolumn{1}{c}{$\rm{[km]}$} &
    \multicolumn{1}{c}{$\rm{[M_\odot]}$} & 
    \multicolumn{1}{c}{$\rm{[Myr]}$}
    \\
    \hline  
    $167$ & $20.4$ & $21.38$ & $9.98$ & $15$ & $9.55\pm0.11$ & $26,400\pm200$ & $260\pm21$ & $2,230^{+280}_{-250}$ & $1.317^{+0.014}_{-0.018}$ & $556^{+15}_{-22}$
    \\
    \end{tabular}
\end{table}


\subsection{Theoretical Mass-Radius Relation}\label{sec:massrad}

We gain additional insight into WD1 by comparing our results to theoretical mass-radius relations at different compositions. For this comparison, we first use available Galex \citep{2005ApJ...619L...1M} and Pan-STARRS \citep{2016arXiv161205560C} photometry for an alternate measure of the radius of WD1. To do so, we again employ \cite{2011ApJ...730..128T} DA models. Synthetic spectra are normalized using the distance determined from Gaia EDR3 parallax. Using the normalized synthetic spectra, we compute synthetic photometry using the \texttt{pyphot} package\footnote{https://mfouesneau.github.io/docs/pyphot/} and fit for the radius and $T_{\rm eff}$ using the Levenberg-Marquardt non-linear least squares minimization routine. Note that while SDSS \citep{2017arXiv171103234K} photometry is available for WD1, we found that these data were less self-consistent than Pan-STARRS photometry and opted to leave these data out of our fitting routine. The best fit returns $R=2600\pm200$~km and $T_{\rm eff} = 26,000\pm3,000$~K (right panel of Fig.~\ref{fig:WD1_spect})

\begin{figure}
    \centering
    \includegraphics[width=0.8\textwidth]{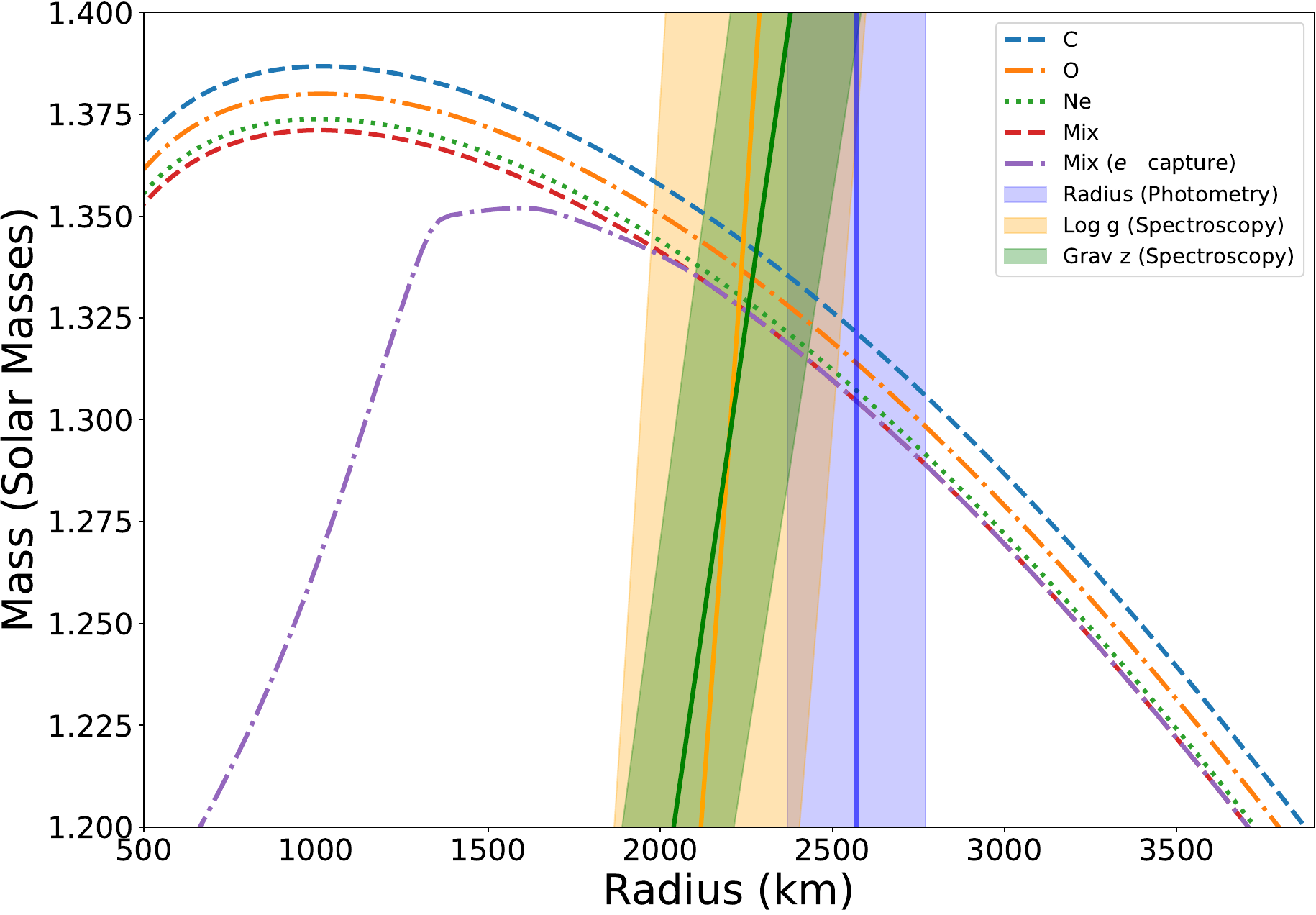}
    \caption{Mass-radius relation at the approximate temperature of WD1 ($26,000$~K) for different compositions. The purple dashed curve is the expected core composition of WD1 formed from the expected mixture due to carbon burning including the effects of electron capture. Solid blue band shows the best fit and 1$\sigma$ region for the radius obtained from photometry, while the orange and green bands show the same for the surface gravity and gravitational redshift determined from spectroscopy.} 
    \label{fig:WD1_mass_rad_relation}
\end{figure}

We utilize theoretical mass-radius relation compositions as calculated and described in \cite{2021Natur.595...39C}, which consider general relativistic corrections to WD structure that are essential at this high mass. These include homogeneous carbon, oxygen, and neon compositions, along with the expected carbon-burning mixture of $58\%$ oxygen, $30\%$ neon, $5\%$ magnesium, $5\%$ sodium, and $2\%$ carbon \citep{2019A&A...625A..87C}. WDs with this composition have a maximum stable mass of approximately 1.369~M$_\odot$ \citep{2022A&A...668A..58A}. We also include relativistic models that include electron capture on sodium, magnesium, and neon, which reduce the maximum allowed mass to approximately 1.35~M$_\odot$ \citep{2021Natur.595...39C}. For this composition (dashed purple curve in Fig.~\ref{fig:WD1_mass_rad_relation}), the mass-radius relation derived from the Balmer-line fitting, the derived gravitational redshift, and the derived photometric radius agree within uncertainties.

\section{Alternate Cluster Age}\label{sec:alt_cluster_age}

The determination of the age of the Hyades cluster from isochrones depends sensitively on the assumed reddening and metallicity and is typically driven by a handful of main sequence turnoff and giant stars. This is also true for age derivations from known cluster member white dwarfs \citep{2009ApJ...696...12D,2019A&A...623A..35L} due to the lengthy progenitor lifetimes determined from the same isochrone models. Given the high mass of WD1 and its expected short progenitor lifetime, we employ WD1 as an alternate means of constraining the cluster's age which does not depend sensitively on isochrones.

In recent work on the IFMR, \cite{2022ApJ...926L..24M} identified and characterized a 
$1.20\pm0.01$~M$_\odot$ WD with a cooling age of $45\pm4$~Myr born in the $81\pm6$~Myr Alpha Persei cluster (Gaia EDR3 439597809786357248, hereafter AP1). This work is relevant because the cluster's age was determined kinematically \citep{2021arXiv211004296H}, without the need to employ isochrone models. While the high-mass region of the IFMR needs to be better developed for a confident progenitor mass or lifetime estimate for WD1, we can reasonably expect that it came from a precursor that was more massive than AP1. Thus, we expect a precursor lifetime less than the $36\pm7$~Myrs lifetime estimate for the progenitor of AP1. This allows us to place an upper bound on the progenitor lifetime of WD1 of $36$~Myrs, leading to a WD age constraint on the Hyades cluster of $<606$~Myrs. Though this does not account for the error in the progenitor lifetime of AP1, given the difference in mass between WD1 and AP1 of $>0.1$~M$_\odot$, we do not expect the progenitor lifetime of WD1 to lie above the mean value for AP1. This alternate cluster age estimate depends on two cooling age estimates from spectroscopy, and the kinematic age of Alpha Persei, without relying on cluster isochrones.


\section{Conclusions}\label{sec:conclusions} 

In this contribution, we examined the apparent deficiency of white dwarfs in the nearby Hyades cluster by employing a technique that attempts to reconstruct open clusters by identifying stars that may have escaped from their environs. From this search, we identified three candidate ultramassive white dwarfs with kinematics consistent with having escaped the cluster. Two of the three are in a mass range that inhibits confident cluster association, while the remaining object appears to be a high-probability escapee. For this high-probability escapee, we obtained follow-up spectroscopy with Gemini-North GMOS.

The spectrum reveals features indicative of a hydrogen atmosphere white dwarf likely formed via single-stellar evolution. From non-local thermodynamic equilibrium white dwarf atmospheric models and oxygen-neon core cooling models, we estimate a mass of $1.317^{+0.014}_{-0.018}$~M$_\odot$, a cooling age of $556^{+15}_{-22}$~Myr, and a radius of $2,230^{+280}_{-250}$~km. The derived mass is amongst the highest known for any white dwarf, particularly for those that appear to be single stellar evolutionary products. This provides a critical observational benchmark for white dwarfs created from single progenitor stars, demonstrating that single stars can produce white dwarfs with masses close to the Chandrasekhar limit. 

It is interesting that such a high-mass white dwarf was identified as having been born in the Hyades cluster. The Hyades is not exceptionally rich in stars nor in a particularly dense region of the Galaxy; by most accounts, it is a typical moderately populated and evolved cluster. The sole factor that makes the cluster stand out is its proximity as the closest cluster to the Sun. This enables the detection of older, cooler white dwarfs and the ability to trace back escaped stars with greater precision, allowing us to study the cluster in greater detail than any other. The combination of the unremarkable nature of the Hyades cluster and the benefits of its proximity suggests that open star clusters may be producing ultramassive white dwarfs, including white dwarfs which push the Chandrasekhar limit, more commonly than previously thought.

\begin{acknowledgments}

The authors would like to thank Chris Matzner for valuable discussions, and Pier-Emmanuel Tremblay for providing atmospheric models extended to $\log g$ of $10$~cm~s~$^{-2}$. This research has made use of the SIMBAD and Vizier databases, operated at CDS, Strasbourg, France and the Montreal White Dwarf Database produced and maintained by Prof. Patrick Dufour (Université de Montrèal) and Dr. Simon Blouin (University of Victoria). This work was supported by the Natural Sciences and Engineering Research Council of Canada. This work includes results based on observations obtained at the international Gemini Observatory, a program of NSF’s NOIRLab, which is managed by the Association of Universities for Research in Astronomy (AURA) under a cooperative agreement with the National Science Foundation on behalf of the Gemini Observatory partnership: the National Science Foundation (United States), National Research Council (Canada), Agencia Nacional de Investigaci\'{o}n y Desarrollo (Chile), Ministerio de Ciencia, Tecnolog\'{i}a e Innovaci\'{o}n (Argentina), Minist\'{e}rio da Ci\^{e}ncia, Tecnologia, Inova\c{c}\~{o}es e Comunica\c{c}\~{o}es (Brazil), and Korea Astronomy and Space Science Institute (Republic of Korea). This work has made use of data from the European Space Agency (ESA) mission {\it Gaia} (\url{https://www.cosmos.esa.int/gaia)}, processed by the {\it Gaia} Data Processing and Analysis Consortium (DPAC, \url{https://www.cosmos.esa.int/web/gaia/dpac/consortium}). Funding for the DPAC has been provided by national institutions, in particular the institutions participating in the {\it Gaia} Multilateral Agreement. The Pan-STARRS1 Surveys (PS1) and the PS1 public science archive have been made possible through contributions by the Institute for Astronomy, the University of Hawaii, the Pan-STARRS Project Office, the Max-Planck Society and its participating institutes, the Max Planck Institute for Astronomy, Heidelberg and the Max Planck Institute for Extraterrestrial Physics, Garching, The Johns Hopkins University, Durham University, the University of Edinburgh, the Queen's University Belfast, the Harvard-Smithsonian Center for Astrophysics, the Las Cumbres Observatory Global Telescope Network Incorporated, the National Central University of Taiwan, the Space Telescope Science Institute, the National Aeronautics and Space Administration under Grant No. NNX08AR22G issued through the Planetary Science Division of the NASA Science Mission Directorate, the National Science Foundation Grant No. AST-1238877, the University of Maryland, Eotvos Lorand University (ELTE), the Los Alamos National Laboratory, and the Gordon and Betty Moore Foundation. Funding for the Sloan Digital Sky Survey V has been provided by the Alfred P. Sloan Foundation, the Heising-Simons Foundation, the National Science Foundation, and the Participating Institutions. SDSS acknowledges support and resources from the Center for High-Performance Computing at the University of Utah. The SDSS web site is \url{www.sdss.org}. SDSS is managed by the Astrophysical Research Consortium for the Participating Institutions of the SDSS Collaboration, including the Carnegie Institution for Science, Chilean National Time Allocation Committee (CNTAC) ratified researchers, the Gotham Participation Group, Harvard University, Heidelberg University, The Johns Hopkins University, L’Ecole polytechnique f\'{e}d\'{e}rale de Lausanne (EPFL), Leibniz-Institut f$\ddot{\rm u}$r Astrophysik Potsdam (AIP), Max-Planck-Institut f$\ddot{\rm u}$r Astronomie (MPIA Heidelberg), Max-Planck-Institut f$\ddot{\rm u}$r Extraterrestrische Physik (MPE), Nanjing University, National Astronomical Observatories of China (NAOC), New Mexico State University, The Ohio State University, Pennsylvania State University, Smithsonian Astrophysical Observatory, Space Telescope Science Institute (STScI), the Stellar Astrophysics Participation Group, Universidad Nacional Aut\'{o}noma de M\'{e}xico, University of Arizona, University of Colorado Boulder, University of Illinois at Urbana-Champaign, University of Toronto, University of Utah, University of Virginia, Yale University, and Yunnan University. This work includes data collected by the Galex mission which is publicly available from the Mikulski Archive for Space Telescopes (MAST, https://archive.stsci.edu/). This work made use of Astropy:\footnote{http://www.astropy.org} a community-developed core Python package and an ecosystem of tools and resources for astronomy \citep{astropy:2013, astropy:2018, astropy:2022}. Gemini spectra were processed using the Gemini IRAF package.
\end{acknowledgments}

\vspace{5mm}
\facilities{Gaia (DR2 \& EDR3), Gemini-North (GMOS).}

\software{Astropy \citep{astropy:2013, astropy:2018, astropy:2022}, pyphot.}

\section*{Data Availability}

We constructed the cluster member and escapee catalogues from the Gaia EDR3 database. Interloper analysis made additional use of the Fusillo Gaia EDR3 WD catalogue. Data from SDSS, GALEX, and Pan-STARRS1 were used in photometric and preliminary analysis. Each of the aforemention catalogues are publicly available.


\bibliography{main}{}
\bibliographystyle{aasjournal}


\end{document}